\documentclass[]{aa}
\usepackage[dvips]{graphicx}
\usepackage{txfonts}
\usepackage{psfig}
\begin{document}
\title{SMA observations of the proto brown dwarf candidate \\
SSTB213 J041757}

   \author{N. Phan-Bao
       \inst{1,2}
       \and
       C.-F. Lee
       \inst{2}
       \and
       P.T.P. Ho
       \inst{2,4}
       \and
       E.L. Mart\'{\i}n
       \inst{3}
       }

\offprints{N.~Phan-Bao}

\institute{Department of Physics, HCM International University-VNU, Block 6, Linh Trung Ward, Thu Duc
           District, HCM, Viet Nam. \\
           \email{pbngoc@hcmiu.edu.vn}
          \and
          Institute of Astronomy and Astrophysics, Academia Sinica. 
          PO Box 23-141, Taipei 106, Taiwan, ROC. 
          \and
          Centro de Astrobiologia (CAB-CSIC), Ctra. Ajalvir km 4, 28850 Torrej\'on de Ardoz, Madrid, Spain.        
          \and
	  Harvard-Smithsonian Center for Astrophysics, Cambridge, MA, USA.
    	  }

      \date{Received; accepted}

\abstract
{The previously identified source SSTB213 J041757 is a proto brown dwarf candidate in Taurus, 
which has two possible components A and B. 
It was found that component B is probably a class 0/I proto brown dwarf 
associated with an extended envelope.}
{Studying molecular outflows from young brown dwarfs provides important insight into brown dwarf
formation mechanisms, particularly brown dwarfs at the earliest stages such as class 0, I.
We therefore conducted a search for molecular outflows from SSTB213 J041757.}
{We observed SSTB213 J041757 with the Submillimeter Array to search for CO molecular outflow emission from
the source.}
{Our CO maps do not show any outflow emission from the proto brown dwarf candidate.}
{The non-detection implies that the molecular outflows from the source are weak; 
deeper observations are therefore needed to probe the outflows from the source.}

\authorrunning{Phan-Bao et al.}
\titlerunning{SSTB213 J041757}

  \maketitle

\section{Introduction}
Since molecular outflows are a basic component of the star formation process,
studying the molecular outflow properties will help us
understand the star formation mechanism. For brown dwarfs (BD),
some observations have been carried out in the past few years to 
characterize jets and molecular outflows in class II BDs, very low-mass (VLM)
stars (Whelan et al. \cite{whelan05}, Phan-Bao et al. \cite{pb08,pb11}), 
and proto BD candidates (Bourke et al. \cite{bourke}, Kauffmann et al. \cite{kau}).
These observations have suggested that the outflow process occurs in BDs
as a scaled-down version of that in low-mass stars, providing additional evidence
that BDs form like stars. 
However, it is still unclear how this physical process occurs at earlier stages,
such as at classes 0, I of the BD formation process, 
because we lack identification and studies of BDs at these classes.
Therefore, observations of molecular outflows from BDs at these earliest stages 
are clearly important to complete our understanding of BD formation mechanism.
These observations will also provide strong constraints 
on BD formation theory (e.g., Machida et al. \cite{machida}).

Taurus is a nearby star-forming region ($\sim$145 pc) where many
class II BDs have been identified and studied. The region is therefore a good target
to study the BD formation process.
For class II BDs and VLM stars, Phan-Bao et al. (\cite{pb11}) have reported the first detection of
a bipolar molecular outflow from MHO~5, a VLM star of 90~$M_{\rm J}$.
The molecular outflow from MHO~5 shows similar properties as seen in ISO-Oph~102 (a young BD in $\rho$
Ophiuchi, Phan-Bao et al. \cite{pb08}) 
such as low-velocity ($<$5~km~s$^{-1}$), compact structure (500$-$1000~AU), 
small outflow mass ($10^{-6}-10^{-3}M_{\odot}$), and 
low mass-loss rate ($10^{-10}-10^{-6}M_{\odot}yr^{-1}$).
For BDs at ealier classes, Barrado et al. (\cite{barrado}) have reported
the detection of a proto BD candidate SSTB213 J041757 (hereafter J041757) 
with two possible components, SSTB213 J041757 A and SSTB213 J041757 B 
(hereafter J041757-A and J041757-B).
Luhman et al. (\cite{luhman10}) spectroscopically classified 
J041757-A as an M2 background dwarf star and suggested that the proper motion
of J041757-B is inconsistent with membership in Taurus.
Palau et al. (\cite{palau}) recently reported a detection of centimeter continuum emission
at the position of J041757-B, which is attributed to thermal free-free emission
due to shocks in the jet of J041757-B driven by a central object. 
The detection has implied that J041757-B might be a proto BD. 
J041757-B is therefore a good target for our ongoing program of characterizing molecular outflows in the 
substellar domain. We thus conducted a search for molecular outflows from 
the proto BD candidate with the Submillimeter Array (SMA).
This paper presents our millimeter observations of J041757 and discusses the nature
of the source. 
\begin{table*}
  \caption{SMA observing log for SSTB213 J041757}
\label{log}
  $$
 \begin{tabular}{llccc}
   \hline 
   \hline
   \noalign{\smallskip}
Target   & ~~~~~~~~~~~~~~~~Position                         &  Configuration & Beam size                  & 1.3~mm continuum emission  \\
         & $\alpha$(J2000)~~~~~~~~~~~$\delta$(J2000) &                & ($\arcsec \times \arcsec$) & (mJy) \\
J041757  & 04$^{\rm h}$ 17$^{\rm m}$ 57.75$^{\rm s}$ +27$^{\rm o}$ 41$\arcmin$ 05.5$\arcsec$          
&   Compact      & $3.34 \times 2.79$           &  $<$1  \\
   \hline
   \end{tabular}
   $$
\end{table*}

\section{Observation and data reduction}
We observed J041757 at 230~GHz with SMA \footnote{
The Submillimeter Array is a joint project between the 
Smithsonian Astrophysical Observatory and 
the Academia Sinica Institute of Astronomy and Astrophysics 
and is funded by the Smithsonian Institution and the Academia Sinica.} (Ho et al. \cite{ho})
on 2012 November 21.
However, the weather was bad because zenith opacities at 225 GHz were above 0.4
during the observations.
Therefore, we re-observed the target on 2012 November 24
with zenith opacities at 225 GHz in the range 0.12-0.26.
The observing log is given in Table~\ref{log}.
The two 4 GHz-wide sidebands, which are separated
by 8 GHz, were used. The SMA correlator was configured with high spectral
resolution bands of 512 channels per chunk of 104~MHz for $^{12}$CO, $^{13}$CO, 
and C$^{18}$O~$J=2-1$ lines, giving a channel spacing of 0.27~km~s$^{-1}$.
A lower resolution of 3.25~MHz per channel was set up for the remainder of
each sideband. The quasars 3C~111 and 3C~84 
were observed for gain calibration and 3C~279 for passband calibration.
Uranus was used for flux calibration of the target. 
The uncertainty in the absolute flux calibration is about 10\%. 

We used the MIR software package and 
the MIRIAD package adapted for the SMA for data calibration and analysis, respectively. 
All eight antennas in the compact 
configuration results in a synthesized beam of 
3$''$.34~$\times$~2$''$.79 with a position angle of 30$^{\circ}$ 
using the natural weighting. The primary FWHM  
beam is about 50$''$~at the observed frequencies. 
The rms sensitivity was about 1~mJy for the continuum,
using both sidebands, and $\sim$0.15~Jy~beam$^{-1}$ per channel for
the line data. 
\begin{figure*}
\psfig{width=16.0cm,file=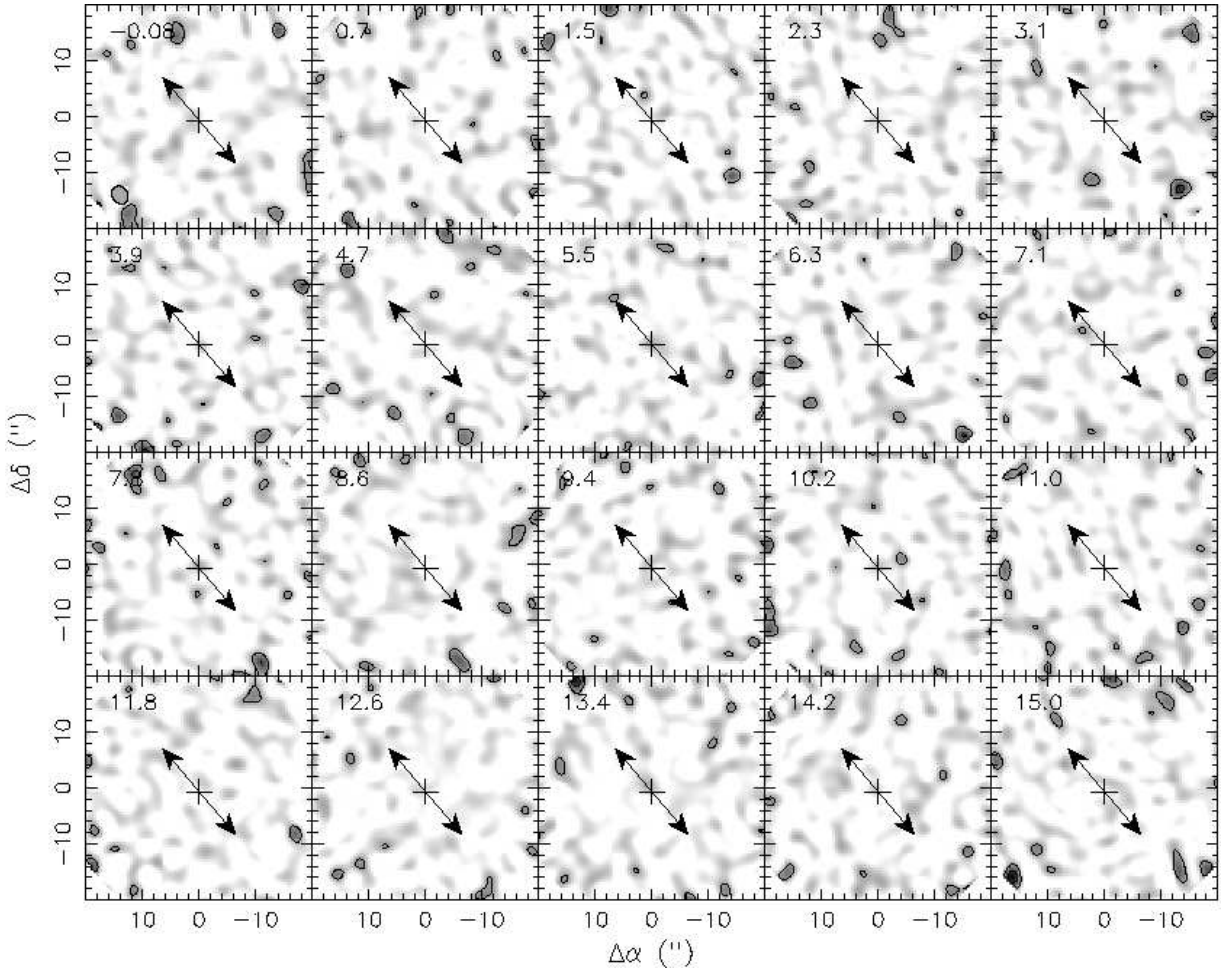,angle=-90}
\caption{Velocity channel maps of the $^{12}$CO~$J=2-1$ emission in the region of 
the proto BD candidate J041757-B, from $\sim$0 to 15~km~s$^{-1}$ with a spacing of 0.8~km~s$^{-1}$.
The position of J041757-B is indicated by the cross mark.
The LSR velocity is indicated in the upper left corner of each panel. 
The contour level starts from 2~$\sigma$, with $\sigma = 0.09$~Jy/beam.
The arrow indicates the expected outflow direction of about 40$\degr$, as suggested
in Palau et al. (\cite{palau}). 
The synthesized beam is shown in the bottom left corner of the first panel.  
\label{spectra_redobj}}
\end{figure*}
\begin{figure*}
\psfig{width=18.0cm,file=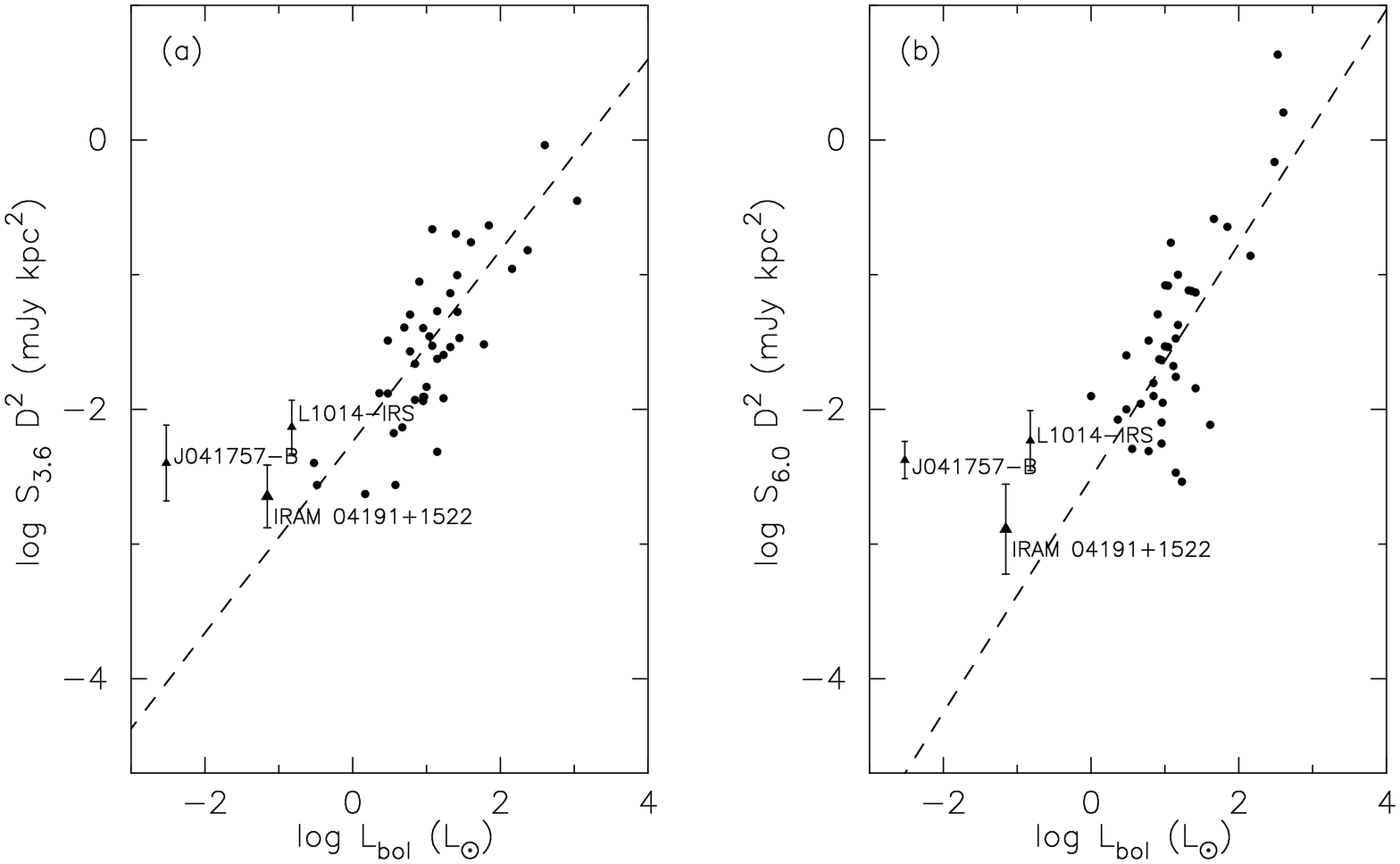,angle=-90}
\caption{3.6 and 6.0 cm luminosity vs. bolometric luminosity for proto stars, data from
Furuya et al. (\cite{furuya}), Eiroa et al. (\cite{eiroa}) and references therein.
Three proto BD candidates IRAM 04191+1522, L1014-IRS and J041757-B are also plotted
with centimeter flux from Andr\'e et al. (\cite{andre99}), Shirley et al. (\cite{shirley}),
Palau et al. (\cite{palau}); the distance and bolometric luminosity for IRAM 04191+1522 and L1014-IRS
are taken from Maheswar et al. (\cite{maheswar}), an average distance of 145$\pm$5~pc is used for 
J041757-B (e.g., Loinard et al. \cite{loinard}). 
The dashed line represents the correlations given in Shirley et al. (\cite{shirley}). 
\label{flux_lbol}}
\end{figure*}
\section{Discussion}
Based on IRAC/Spitzer color-color and color-magnitude diagrams,
Barrado et al. (\cite{barrado}) classified
J041757 as a class 0/I proto BD candidate because
the position of J041757 lies in the locus of class 0/I
objects in the diagrams. Optical and near-infrared images have shown
that J041757 has two possible components A and B. 
The resemblance in spectral energy distribution (SED) between
J041757 and the very low luminosity object L1014-IRS (Young et al. \cite{young},
Bourke et al. \cite{bourke}), which is a class 0/I
proto BD candidate, suggests that J041757 is also a class 0/I proto BD candidate
(see Fig.~3 in Barrado et al. \cite{barrado}).

Using near-infrared spectroscopy, Luhman \& Mamajek (\cite{luhman10})
classified J041757-A as an M2 background dwarf star because of the absence
of strong H$_{2}$O absorption bands in its spectrum as seen in young late-M dwarfs.
For the remaining component J041757-B, the source was too weak to obtain a spectrum.

Barrado et al. (\cite{barrado}) detected unresolved radio continuum emission with VLA
at 6 cm from the position of J041757. 
With higher angular resolution observations, 
Palau et al. (\cite{palau}) recently reported radio continuum emission detected
at 3.6 cm (0.19$\pm$0.11~mJy) and 6 cm (0.20$\pm$0.05~mJy), and the emission is only 
associated with J041757-B; 
J041757-A shows no radio emission. Palau and collaborators also found that
the emission is not polarized with a flat spectral index 
in the range of 6$-$3.6 cm. This suggests that the mechanism of centimeter
continuum emission is thermal free-free emission caused by shocks in the jet
driven by the central object (see Palau et al. \cite{palau} and references therein).
The authors concluded that the emission is not consistent with the centimeter emission from AGNs,
which is typically produced by synchrotron emission (from relativistic accelerated electrons).

If J041757-B is driving a radio jet, molecular outflows,
which are ambient gas swept up by the jet, are expected. 
The Palau et al. (\cite{palau}) observations of $^{12}$CO~$J=1-0$ and $^{13}$CO~$J=1-0$
(see their Fig.~4) suggested that molecular outflow emission from J041757-B
would be detectable in the range of 4.5$-$7.5~km~s$^{-1}$.
However, our observations do not reveal any outflow emission from the region of J041757-B.
Figure~1 presents the velocity channel maps of $^{12}$CO~$J=2-1$ emission in the region of 
J041757-B. No CO outflow emission is detected in the velocity range of 0$-$15~km~s$^{-1}$.
In addition, no $^{13}$CO~$J=2-1$ and C$^{18}$O~$J=2-1$ 
are detected in the same region (see Figs.~A.1 and A.2).
These non-detections indicate two possible scenarios:
\begin{enumerate}
\item We consider the first possible scenario that J041757-B is a background object behind Taurus.
The flat spectral index of radio emission as reported in Palau et al. (\cite{palau})
supports the scenario that the source is a galactic object and not a galaxy.
To estimate the distance of J041757-B, we used the extinction law as given in Mathis (\cite{mathis})
with $R_{\rm V} = 3.1$ and then dereddened apparent magnitudes from Barrado et al. (\cite{barrado}).
Dereddenning the $i$ and $J$-band magnitudes of J041757-A with $A_{\rm V} = 5$
gives a color index $i-J = 2.2$, in good agreement with the color of an M2 dwarf
(West et al. \cite{west}). 
At this point, we derive a distance of about 670~pc for J041757-A.
Our $A_{\rm V}$ estimate is consistent with the Palau et al. values.
With $A_{\rm V} = 5$, 
we obtain dereddened magnitudes, $i = 19.81$, $J = 17.74$, $H = 17.16$ and 
$K_{\rm S} = 16.26$ for J041757-B.
The $i$ and $z$-band magnitudes for J041757-A measured with MegaCam/CFHT
(Barrado et al. \cite{barrado}) agree well with those from SDSS (DR9),
suggesting that the $iz$ photometry of MegaCam/CFHT and SDSS is similar
(as also pointed out in Luhman \& Mamajek \cite{luhman10}).
We therefore used color transformations between SDSS and Johnson-Cousins
as given in Jordi et al. (\cite{jordi}) to calculate $I$-Cousins magnitude
for J041757-B and derive $I_{\rm C}=19.20$.  
The color $I_{\rm C}-J = 1.46$ is consistent with an M4-M5 red giant
(e.g., Bessell \& Brett \cite{bessell88}). 
Comparison of the $J$, $H$ and $K$-band magnitudes of J041757-B
with the $J$, $H$, $K$-band absolute magnitudes of an M4 giant (Th\'e et al. \cite{the}) 
gives an average distance of about 340~kpc for J041757-B, significantly larger than the
currently known extent of our Galaxy as discussed in Palau et al. (\cite{palau}).

The source J041757-B might be a variable giant star, thus its photometric 
variability along with high extinction might cause a large uncertainty in estimating distances
and spectral type. 
However, the $K$-band magnitudes measured at three different epochs do not show any 
significant variability of the source:
$K = 16.95\pm0.08$ (UKIDSS-DR6\footnote{http://vizier.u-strasbg.fr/viz-bin/VizieR}, 
epoch = 2005.968), $K = 16.79\pm0.04$ 
(UKIDSS-DR8, epoch = 2010.077) 
and $K_{\rm S} = 16.81\pm0.03$ (Barrado et al. \cite{barrado}, epoch = 2007.844). 
All these measurements agree well within error bars and 
the difference of about 0.1~mag between the WFCAM and 2MASS photometric systems
(Hewett et al. \cite{hew}).
\item The second possible scenario is that 
molecular outflows from J041757-B are too weak to be detectable with SMA.
Palau et al. (\cite{palau}) reported a detection of blueshifted emission at velocities in the
range 2$-$4~km~s$^{-1}$ for $^{12}$CO~$J=1-0$ (see their Figs.~4 and 5),
which possibly comes from an outflow.
From Figures~4 and 5 of Palau and collaborators, we estimate
its brightness temperature to be about 1~K or $\sim$5.3~Jy/beam (for a beam size of 22$''$).
This corresponds to 0.09 Jy/beam (for an SMA beam size of 3$''$), giving a detection level of 
only $\sim$1$\sigma$ for our $^{12}$CO~$J=2-1$ observations in the same velocity range 
if we assume that the CO~$J=2-1$/$J=1-0$ ratio is 1.
The detection level of $^{13}$CO~$J=2-1$ outflows were estimated 
in the same manner. The brightness temperature of redshifted $^{13}$CO~$J=1-0$ 
emission (e.g., 6.5$-$7.5~km~s$^{-1}$, see Fig.~5 of
Palau et al. \cite{palau}) is about 3~K or $\sim$12.0~Jy/beam (for a beam size of 22$''$).
This corresponds to 0.22 Jy/beam (for an SMA beam size of 3$''$), giving a detection level 
of $\sim$2$\sigma$ for our $^{13}$CO~$J=2-1$ observations.
This therefore could explain our non-detections of $^{12}$CO~$J=2-1$ and $^{13}$CO~$J=2-1$ 
emission.

Moreover, J041757-B and L1014-IRS share similar SEDs
but the latter source shows outflow emission detected by SMA (Bourke et al. \cite{bourke}).
First, this is possibly because the gas in the Taurus region around J041757-B
is much less dense (i.e., weaker outflow emission) than that around L1014-IRS. 
The estimated visual extinction $A_{\rm V}\sim25$ of L1014-IRS (Huard et al. \cite{huard}), which
is significantly greater than the extinction $A_{\rm V}\sim5$ of J041757-B (see above),
supports this possibility. 
Second, the bolometric luminosity of J041757-B 
(0.003~$L_{\odot}$, Barrado et al. \cite{barrado})
is 50 times less luminous than that of L1014-IRS 
(0.15~$L_{\odot}$, 258~pc, Maheswar et al. \cite{maheswar}).
If we assume that the outflow force vs. bolometric luminosity
correlation of proto stars (e.g., Takahashi \& Ho \cite{taka}) 
is applicable for proto BDs, this implies that 
the outflow force in J041757-B would be much weaker (i.e., weaker molecular outflow emission)
than in L1014-IRS.

With such a very low bolometric luminosity (0.003~$L_{\odot}$), 
the centimeter luminosities of J041757-B appear unusual, however.
Figure~2 shows the 3.6 and 6.0 cm luminosity vs. bolometric luminosity diagram
for proto stars ($L_{\rm bol}$ $\leq$ $\sim10^{3}L_{\odot}$, low-mass and intermediate-mass
protostellar sources, see also Shirley et al. \cite{shirley} and references therein)
as well as three proto BD candidates, J041757-B,
L1014-IRS and IRAM~04191+1522 (0.07~~$L_{\odot}$, Maheswar et al. \cite{maheswar}).
The rms dispersions of the proto stars
around the cm luminosity vs. bolometric luminosity correlations
as given in Shirley et al. (\cite{shirley})
are 0.37 and 0.47 for 3.6 cm and 6.0 cm luminosity, respectively.
For L1014-IRS and IRAM~04191+1522, the deviations
from the correlations appear larger than the rms dispersions,
but they are still in the range of dispersion of the proto stars,
0.02$-$0.89 for 3.6 cm and 0.03$-$1.1 for 6.0 cm luminosity.
This suggests that the two proto BD candidates L1014-IRS and 
IRAM~04191+1522 follow the correlations of proto stars.
However, observations of more proto BDs are needed to confirm this trend.
For J041757-B, it is unlikely that the proto BD candidate follows 
the correlations of proto stars.  
With a much lower bolometric luminosity,
the centimeter luminosities of J041757-B, however, are similar to
those of L1014-IRS and IRAM~04191+1522 (Fig.~2). These values of J041757-B
are about 1.6 (for 3.6 cm luminosity) and 2.3 (for 6.0 cm luminosity) 
orders of magnitude brighter than the expected luminosities,
which are estimated from the correlations in Shirley et al. (\cite{shirley}) 
at the given bolometric luminosity of J041757-B. 
This suggests that J041757-B possibly has two components of centimeter emission as
observed in L1014-IRS (Shirley et al. \cite{shirley}), a steady thermal component and a variable
nonthermal component. The contribution of the latter one to the centimeter emission 
could significantly change the luminosities of the proto BD candidate. 
More observations at different epochs are needed to study the variability
of the source. 

Finally, we also measured 1.3~mm continuum emission at the position of J041757-B with an upper limit
of 1~mJy. Our measurement agrees with that from Barrado et al. (\cite{barrado}),
who reported a value of $<$2.88~mJy at 1.2~mm.
Using the relation between the millimeter continuum emission and the mass of compact 
gas and dust as given in Wilking et al. (\cite{wilking}) for a temperature of 20~K,
we derive an upper limit of $\sim$1~M$_{\rm J}$ to the disk mass of J041757-B.
\end{enumerate}

\section{Conclusion}
We presented our SMA observations of the proto BD candidate J041757-B
and discussed two possible scenarios on the nature of the source: a proto BD in Taurus 
and a background giant. 
Based on our observations and currently available data,
we conclude that the molecular outflows from the proto BD candidate 
are weak, therefore, more sensitive radio observations are required to explore
the outflows and the dense envelope/core associated with J041757-B
to confirm its nature.

\begin{acknowledgements}
N.P.-B has been supported by Viet Nam NAFOSTED grant 103.08-2010.17. 
E.M. has been supported by the Spanish Ministry of Economy and Competitiveness (MINECO) 
under grant AyA2011-30147-C03-03 and wishes to thank the Geosciences Department 
at the University of Florida for
a visiting appointment. We also thank the referee for valuable comments.
This work is based in part on data obtained as part of the UKIRT Infrared Deep Sky Survey
and has made use of the Centre de Donn\'ees astronomiques de Strasbourg
(CDS) database.
\end{acknowledgements}

\newpage
\appendix
\section{Velocity channel maps of the $^{13}$CO~$J=2-1$ and C$^{18}$O~$J=2-1$ emission in the
region of J041757-B}

\begin{figure*}
\psfig{width=16.0cm,file=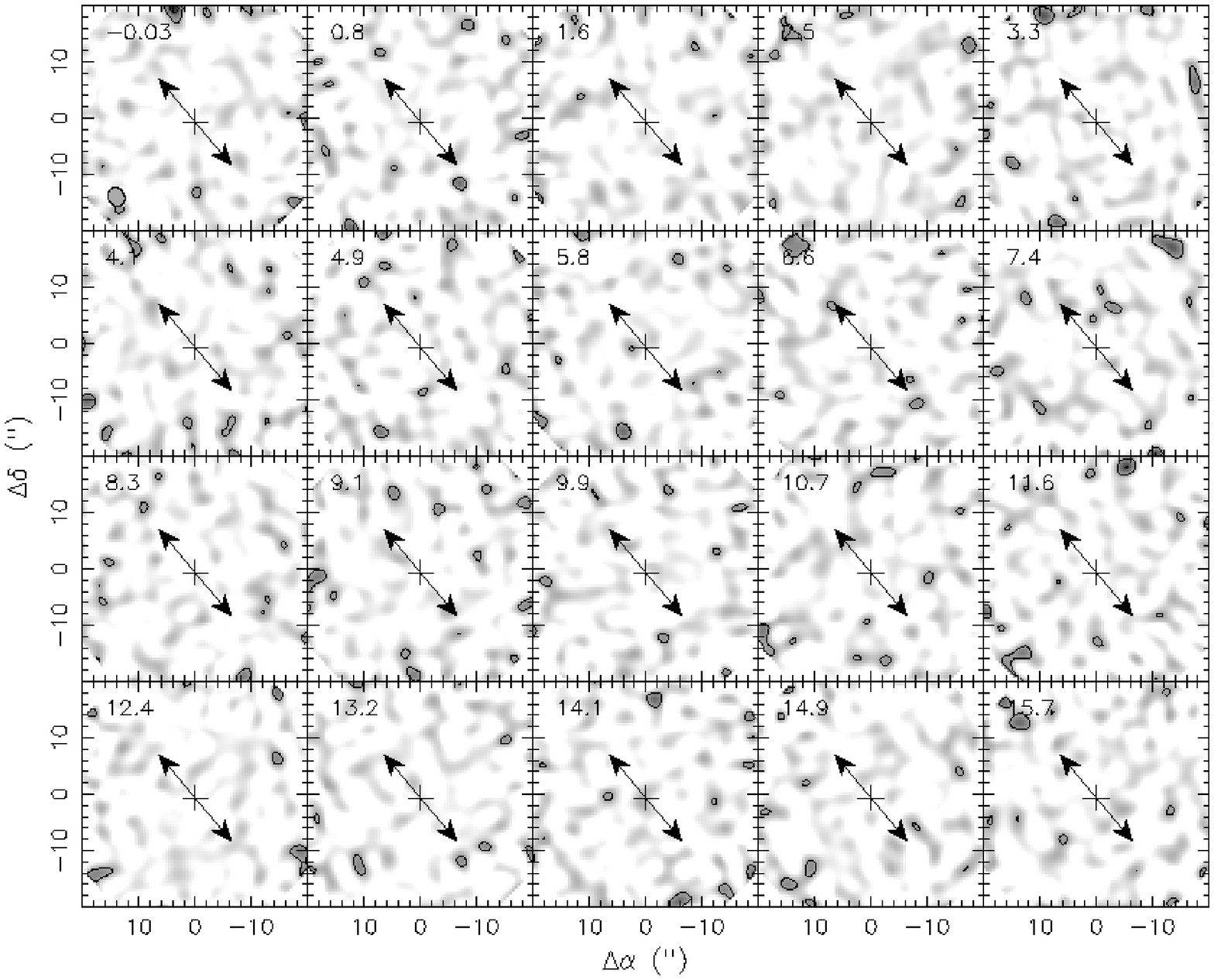,angle=-90}
\caption{Same as Fig.~1 for the $^{13}$CO~$J=2-1$ emission, with $\sigma = 0.085$~Jy/beam.
\label{spectra_redobj}}
\end{figure*}
\begin{figure*}
\psfig{width=16.0cm,file=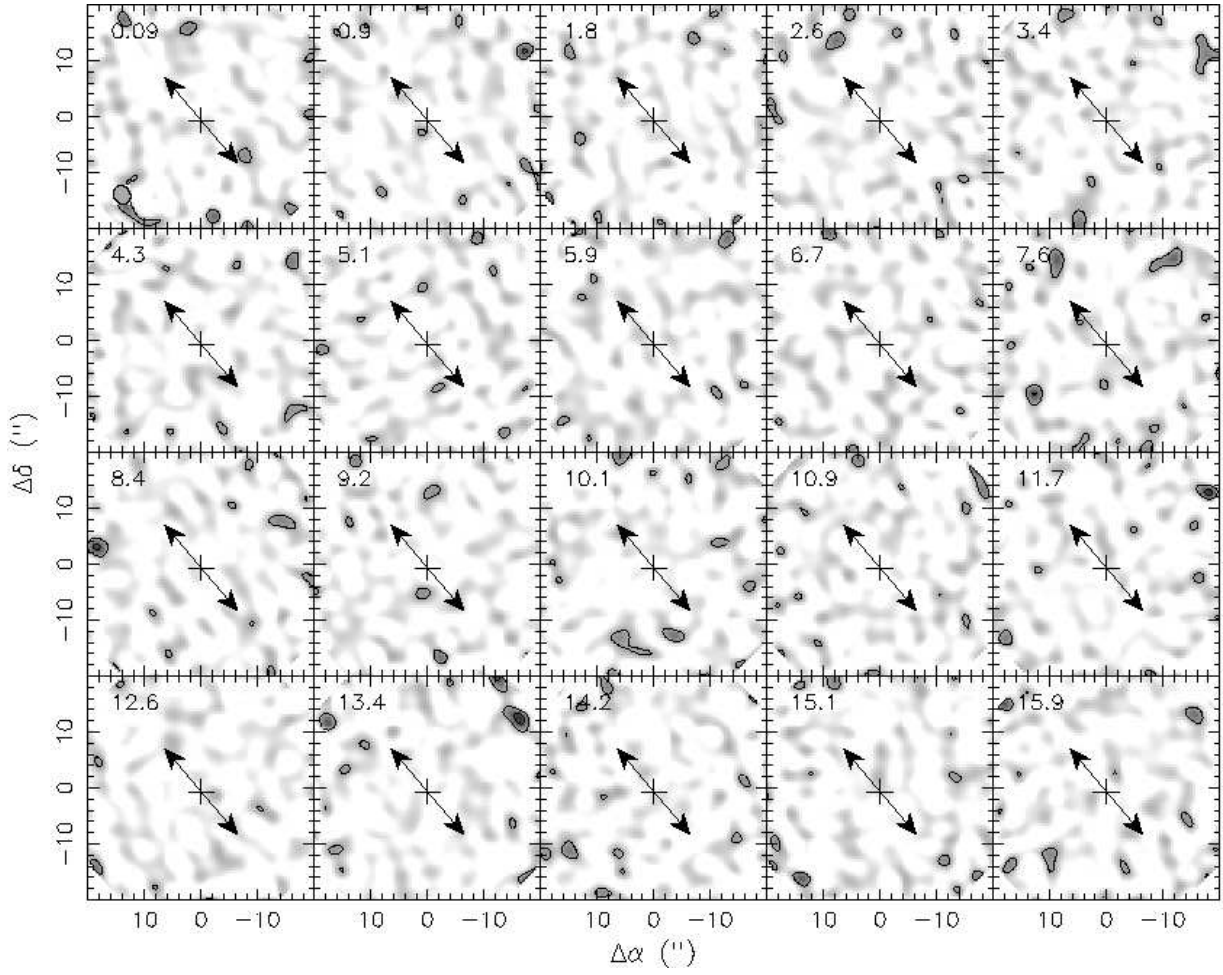,angle=-90}
\caption{Same as Fig.~1 for the C$^{18}$O~$J=2-1$ emission, with $\sigma = 0.083$~Jy/beam.
\label{spectra_redobj}}
\end{figure*}
\end{document}